# Tunable Resonance Coupling in Single Si Nanoparticle-Monolayer WS$_2$ Structures

Sergey Lepeshov[†,$], Mingsong Wang[¶,$], Alex Krasnok[‡,*,$], Oleg Kotov[§], Tianyi Zhang[∥], He Liu[|], Taizhi Jiang[£], Brian Korgel[£], Mauricio Terrones[∥,|,%], Yuebing Zheng[¶,*], and Andrea Alú[‡,*]

† ITMO University, St. Petersburg 197101, Russia

¶ Department of Mechanical Engineering, Texas Materials Institute, The University of Texas at Austin, Austin, TX 78712, USA

‡ Department of Electrical and Computer Engineering, The University of Texas at Austin, TX 78712, USA

§ N.L. Dukhov Research Institute of Automatics, Moscow 127055, Russia

∥ Department of Materials Science and Engineering and Center for 2-Dimensional and Layered Materials, The Pennsylvania State University, University Park, PA 16802, USA

| Department of Chemistry and Center for 2-Dimensional and Layered Materials, The Pennsylvania State University, University Park, PA 16802, USA

% Department of Physics, The Pennsylvania State University, University Park, PA 16802, USA, Department of Materials Science and Engineering & Chemical Engineering, Carlos III University of Madrid, Avenida Universidad 30, 28911 Leganés, Madrid, Spain, and IMDEA Materials Institute, Eric Kandel 2, Getafe, Madrid 28005, Spain

£ McKetta Department of Chemical Engineering and Texas Materials Institute, The University of Texas at Austin, Austin, Texas 78712, USA

$These authors contributed equally.

*E-mail: akrasnok@utexas.edu; alu@mail.utexas.edu; zheng@austin.utexas.edu

## Abstract

Two-dimensional semiconducting transition metal dichalcogenides (TMDCs) are extremely attractive materials for optoelectronic applications in the visible and near-IR range. Here, we address for the first time to the best of our knowledge the issue of resonance coupling in hybrid exciton-polariton structures based on single Si nanoparticles coupled to monolayer

WS$_2$. We predict a transition from weak to strong coupling regime, with a Rabi splitting energy exceeding 200 meV for a Si nanoparticle covered by monolayer WS$_2$ at the magnetic optical Mie resonance. This large transition is achieved due to the symmetry of magnetic dipole Mie mode and by changing the surrounding dielectric material from air to water. The prediction is based on the experimental estimation of TMDC dipole moment variation obtained from measured photoluminescence (PL) spectra of WS$_2$ monolayers in different solvents. An ability of such a system to tune the resonance coupling is realized experimentally for optically resonant spherical Si nanoparticles placed on a WS$_2$ monolayer. The Rabi splitting energy obtained for this scenario increases from 49.6 meV to 86.6 meV after replacing air by water. Our findings pave the way to develop high-efficiency optoelectronic, nanophotonic and quantum optical devices.

Due to the recent tremendous progress in materials science, many novel materials with unique optoelectronic properties have been recently discovered and explored as they exhibit fascinating applications. One prominent example that has been extensively studied in recent years consists of atomically thin semiconducting transition metal dichalcogenides (TMDCs)[1–6]. A monolayer of TMDC is formed by a hexagonal network of transition metal atoms (Mo, W) hosted between two hexagonal lattices of chalcogenide atoms (S, Se). Electronically, TMDCs behave as two-dimensional semiconductors with their bandgaps lying in the visible and near-IR range. In the atomic monolayer limit, these materials are particularly interesting because their bandgap becomes direct, thus enabling enhanced interactions of the dipole transition with light[7]. Due to their monolayer nature, high oscillator strength and the potential for tuning these materials, they become a unique class of 2D materials in the context of optoelectronic applications, such as photodetection and light harvesting[8–11], phototransistors and modulation[12,13], light-emitting diodes[14–16] and lasers[17,18].

A variety of new optical effects stem from the interaction between TMDC monolayers and plasmonic (i.e., made of noble metals) nanoscale objects, which have become a goal of extensive studies. Examples include the observation of strong plasmon-exciton coupling[19–22], pronounced Fano resonance[23], plasmon-induced resonance energy transfer[24] among others. These effects benefit from the small mode value of plasmonic resonances and the strong dipole moment of excitons in TMDCs. However, conventional plasmonic materials, such as gold (Au) and silver (Ag), have finite conductivities at optical frequencies, leading to inherent dissipation of electromagnetic energy. This energy loss causes Joule heating of the structure and its local environment[25,26]. For many applications, heat generation in nanostructures is detrimental, since the behavior of the excitonic system can be modified dramatically with temperature. Moreover, an excitonic system exhibits quenching when it is placed nearby metallic nanostructures due to the dominant non-radiative decay. We also note that plasmonic nanostructures are not compatible with most semiconductor device processing technologies, thus narrowing the scope of their applicability.

High-index semiconductor (e.g., Si, GaP, Ge) nanoparticles can circumvent these issues, while providing an abundant spectrum of optical functions[27–32]. Here, we address for the first time to the best of our knowledge the resonance coupling in single Si nanoparticle-monolayer $WS_2$. The purpose of this paper is twofold. First, to predict a transition from weak to strong coupling for a single Si nanoparticle (SiNP) covered by monolayer $WS_2$ at the magnetic optical Mie resonance. The strong coupling regime is achieved owing to the symmetry of the magnetic Mie mode, which allows a strong interaction with atomically thin materials, and can be largely tuned by changing the surrounding dielectric material. This prediction is based on the experimental estimation of the TMDC dipole moment variation obtained from measured photoluminescence (PL) spectra of $WS_2$ monolayers in different solvents (e.g. air and water). We demonstrate that the core-shell geometry of the hybrid system placed in water may provide a value of the Rabi splitting energy exceeding 200 meV. We also note that a similar core-shell geometry has been recently realized in Refs.[33,34], based on small $SiO_2$ and Au nanoparticles by CVD technology, making a single SiNP-$WS_2$ core-shell design potentially realizable. Second, the ability of such a system to tune the resonance coupling is demonstrated experimentally in the geometry of optically resonant spherical SiNP arranged on a $WS_2$ monolayer. The values of Rabi splitting energy obtained from fitting the experimental data reveal a significant enhancement (from 49.6 meV up to 86.6 meV), when replacing air by water.

We start our analysis by considering a core-shell spherical particle with a silicon (Si) core and a monolayer WS$_2$ shell. This simple model configuration can be analyzed via analytical methods. First, by using Mie theory[35] we calculate separately the scattering cross sections of a Si nanoparticle (NP) and a WS$_2$ empty shell with atomic thickness ($h \approx 0.7$ nm), whose dielectric function is determined by reflectance measurements[36]. The dielectric permittivity of Si has been taken from Ref.[37] for ambient temperature T=300 K. For Si nanoparticles, we obtain strong electric (ED) and magnetic (MD) dipole resonances (Figure 1(a)) whose positions depend on the particle radius. The magnetic quadrupole (MQ) resonance can be also observed in the scattering spectrum (for multipole decomposition see *Supplementary information*). By choosing the radius of Si nanoparticle $R = 75$ nm, we obtain the magnetic dipole resonance at $\lambda \approx 615$ nm (~2 eV), at which the A exciton resonance in WS$_2$ monolayers takes place[36]. The scattering cross section of the WS$_2$ shell is two orders of magnitude less than that of SiNP of same radius (Figure 1(b)), but it demonstrates a sharp peak (Ex), corresponding to the A exciton in WS$_2$.

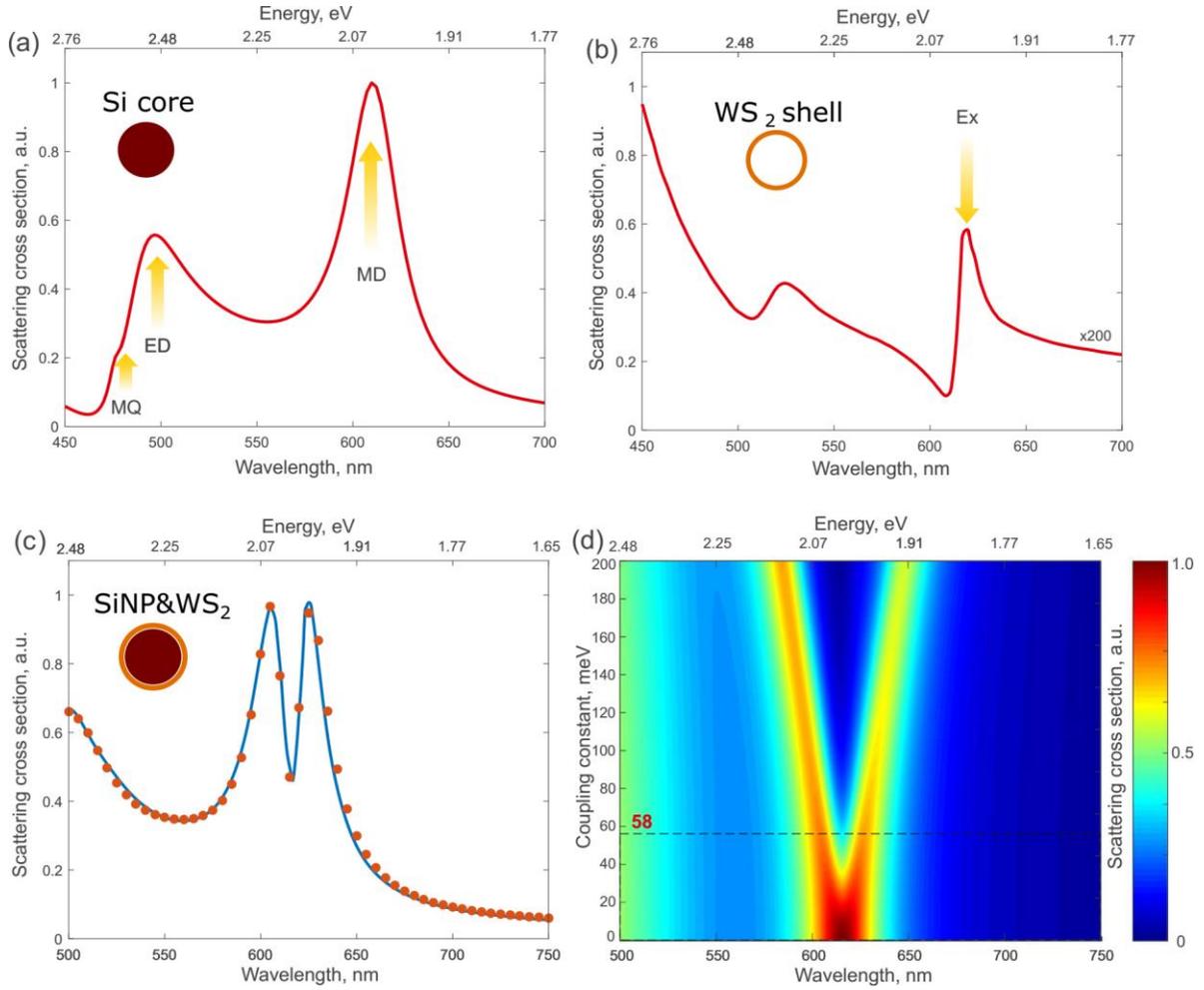

Figure 1: Scattering cross sections of (a) Si nanoparticle of radius R = 75 nm, (b) WS$_2$ shell with thickness $h \approx 0.7$ nm, and (c) Si-WS$_2$ core-shell system with the same parameters. Red dots in (c) shows the CMT fitting. (d) Scattering cross section of Si-WS$_2$ core-shell system at various coupling constants g. The dotted line at g=58 meV corresponds to the case shown in (c).

Then, we use Mie theory generalized for core-shell spherical particles[38–40] for the calculation of the scattering cross section of the SiNP-WS$_2$ core-shell system. We use the parameters fitted earlier to achieve the maximum overlap of the magnetic dipole Mie resonance of SiNP and the A exciton peak of the WS$_2$ monolayer. The interaction of these two resonances gives a strong Fano-like resonance at $\lambda \approx 615$ nm (Figure 1(c)), for which the coupling constant $g$ can be extracted using coupled mode theory (CMT). In the harmonic
5

oscillator approximation (for details see *Supplementary information*), the scattering cross section of our system with two coupled oscillators is given by[23,41,42]:

$$\sigma_{scat} \propto \left| \frac{\omega^2 \tilde{\omega}_{ex}^2}{\tilde{\omega}_{md}^2 \tilde{\omega}_{ex}^2 - g^2 \omega^2} \right|^2, \tag{1}$$

where $\tilde{\omega}_{ex}^2 = \omega^2 - \omega_{ex}^2 + i\gamma_{ex}\omega$ and $\tilde{\omega}_{md}^2 = \omega^2 - \omega_{md}^2 + i\gamma_{md}\omega$ are the harmonic oscillator terms for the exciton and magnetic dipole resonances, respectively, with $\omega_{ex}$, $\omega_{md}$ and $\gamma_{ex}$, $\gamma_{md}$ being the corresponding resonance frequencies and dissipation rates (full width at half maximum). After fitting the curve shown in Figure 1(c), we obtain that the coupling constant $g \approx 58$ meV, whereas the dissipation rates equal $\gamma_{ex} \approx 33$ meV and $\gamma_{md} \approx 84$ meV.

It is known that in order to achieve strong coupling, the system should satisfy the criterion $\hbar\Omega > \hbar\gamma_{ex} + \hbar\gamma_{md}$, where $\hbar\Omega$ is the Rabi splitting energy[19,22]. The coupling strength and Rabi splitting energy are related to each other via $g = \hbar\Omega/2$. In our case, the criterion $\hbar\Omega > \hbar\gamma_{ex} + \hbar\gamma_{md}$ is not satisfied when the system is in air, because the total loss $\hbar\gamma_{ex} + \hbar\gamma_{md}$ is 117 meV, whereas the Rabi splitting energy $\hbar\Omega$ equals 116 meV.

Figure 1(d) demonstrates the dependency of the scattering cross-section on the wavelength and coupling strength $g$ of the SiNP-WS$_2$ core-shell system in air. The dashed line shows the value of the coupling strength in air. The other parameters of the system corresponds to those obtained by our CMT fitting of the scattering spectrum (see Figure 1(c)). This figure shows that strong coupling in such a system can be achieved by increasing $g$.

To enhance the coupling strength, we propose to place the Si-WS$_2$ system in an environment (solvent) with a high static dielectric constant, and utilize the dielectric screening effect[43,44] to increase the exciton dipole moment of the WS$_2$ monolayer. The coupling strength is proportional to the product of the exciton dipole moment $d$, number of excitons $N$ and the tangential component of the electric field near the sphere ( $g \sim d \cdot N \cdot E_t$ )[45–48]. However, an exciton in monolayer WS$_2$ couples with the magnetic dipole resonance in SiNP at high frequencies (visible range), while the screening effect manifests at lower frequencies (depending on the exciton binding energy). Thus, the ideal background material for optimal coupling enhancement should possess a dielectric constant that is small at high frequencies and large at low frequencies. We find that water ideally fits this requirement, since it has $\varepsilon_w \approx 1.77$ in the visible range and $\varepsilon_w \approx 78$ in the static limit. Moreover, in contrast to solids, water due to its liquid state interacts with the WS$_2$ monolayer directly, avoiding air gaps.

Strictly speaking, according to self-consistent many-body theory[49–51], the effective potential of the electron-hole interaction should be taken at the frequency corresponding to the exciton binding energy. In conventional bulk semiconductors such as GaAs, the dielectric constant of the material is rather large ($\sim 13$), which gives a small binding energy ($\sim 5$ meV), and one could approximately use the static dielectric constant to describe the screening effect. However, 2D semiconductors (such as monolayer WS$_2$) in vacuum are essentially unscreened, thus possessing high binding energies ~800 meV. Even for WS$_2$ on a SiO$_2$ substrate it is ~400 meV, lying in the infrared range. Nevertheless, some experiments on the dielectric screening of excitons in TMDC atomically thin monolayers[43,44] demonstrate good agreement with theoretical estimations based on static screening constants.

We estimate the enhancement of the exciton dipole moment in a monolayer WS$_2$ shell of the Si-WS$_2$ core-shell system embedded in water, using the static screening approximation. The Coulomb potential of the electron-hole interaction in TMDC falls off at a distance of $\approx 2$ nm (see Supporting Information in Ref.[43]), so we can treat the 150 nm Si sphere as a flat substrate, and consider the screening in the WS$_2$ monolayer sandwiched between semi-infinite silicon (with static dielectric constant $\varepsilon_s = 11.7$) and water (with $\varepsilon_w = 78$ at $T = 300$ K). In this case, the exciton dipole moment is proportional to the effective dielectric constant $d_w \sim (\varepsilon_s + \varepsilon_w)/2$, while for the system in air it will be $d_a \sim (\varepsilon_s + 1)/2$. Notice that the field lines of an exciton in 2D semiconductors pass mainly through the environment, and thus the screening is almost independent of the dielectric constant of the semiconductor, and it is completely determined by the surrounding media. Thus, assuming that the number of excitons does not change, the required enhancement of the dipole moment of a single exciton can be estimated by $(d_w/d_a)_{\text{theor}} \sim (\varepsilon_s + \varepsilon_w)/(\varepsilon_s + 1) \approx 7$. This estimation is not far from the experimentally obtained value, which gives $(d_w/d_a)_{\text{exp}} \approx 2.7$ (see the discussions of Figure 3(d)). Below we use this experimentally obtained value.

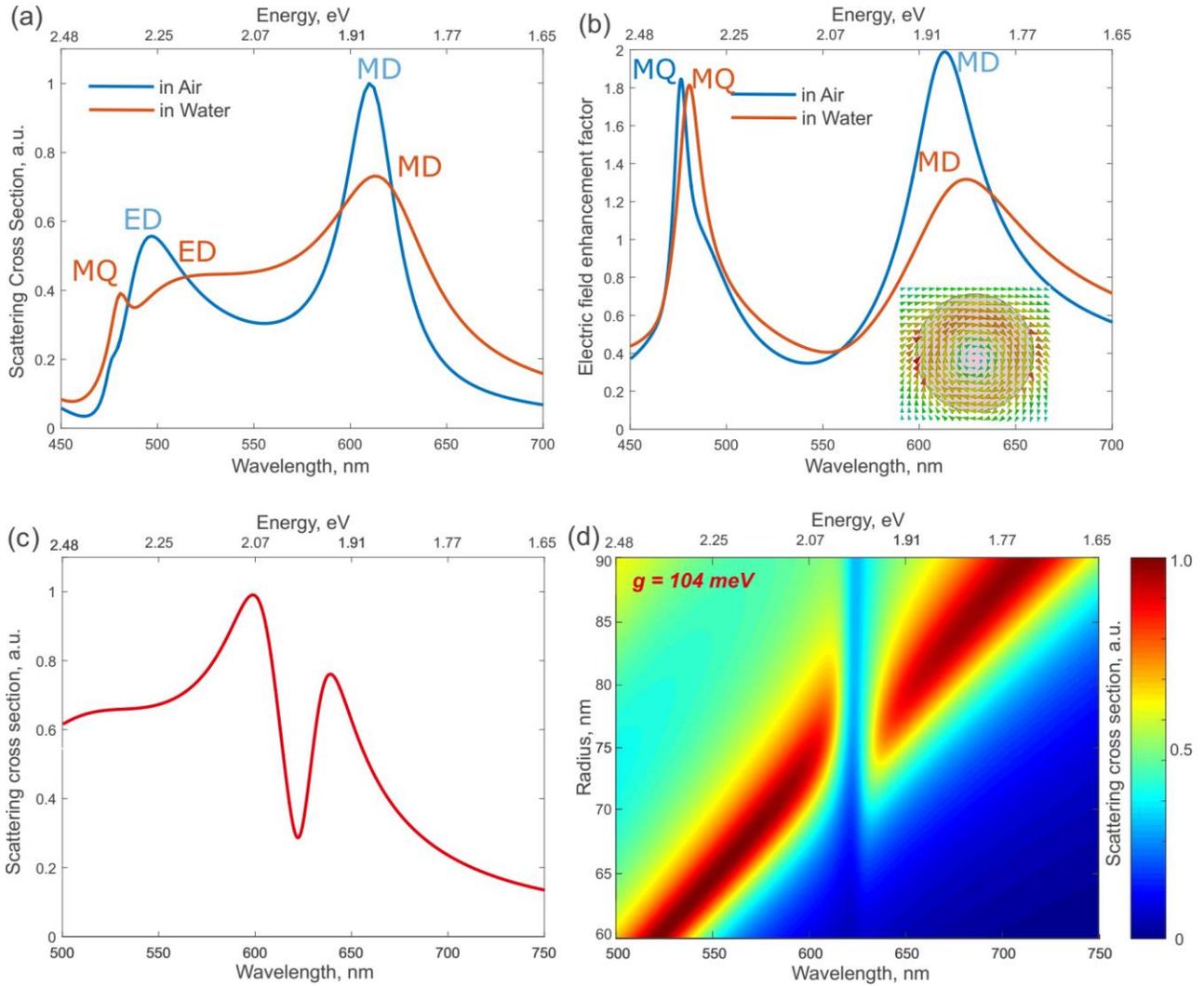

Figure 2: (a) Scattering cross sections of a spherical SiNP with radius R = 75 nm in air and in water environments. (b) Enhancement factor of the tangential electric field near the Si sphere in air and in water environments. Inset shows the electric field at the MD resonance. (c) Scattering cross section of Si-WS$_2$ core-shell system in water. (d) Scattering cross section of Si-WS$_2$ core-shell system in water as a function on wavelength and SiNP radius.

For the estimation of MD shifting, we calculate the scattering cross section of the spherical SiNP embedded in water using Mie theory, and compare it with the previous result in air, as shown in Figure 2(a). For multipole decomposition see *Supplementary information*. We also calculate the enhancement factor of the *tangential electric field* near the SiNP in air (blue curve) and in water (red curve). Figure 2(b) highlights how at the magnetic dipole resonance decreases by ~1.5 times. We note that since the ED mode does not support a tangential electric field near the SiNP, it does not appear in the electric field spectra, see Figure 2(b). The high frequency peak corresponds to the magnetic quadrupole resonance, and it is barely affected by the environment, as it is more confined within the particle. Thus, knowing the enhancement of the dipole moment of the exciton subsystem and the weakening of the tangential electric field of the MD mode, we can estimate the change of coupling constant $g_w/g_a \sim 2.7/1.5 \approx 1.8$ (for the same number of excitons). Then, by obtaining the frequency of the magnetic dipole in the case with water from Figure 2(a), and using CMT with $g = 58 \cdot 1.8 \approx 104$, we plot the scattering cross section of the SiNP-WS$_2$ core-shell system embedded in water (see Figure 2(c)). In this case, the Rabi splitting becomes much more obvious. The coupling constant in this system $g \approx 104$ meV (and Rabi splitting energy 208 meV), whereas the dissipation rates equal $\gamma_{ex} \approx 33$ meV and $\gamma_{md} \approx 173$ meV. Thus, we can conclude that for a SiNP-WS$_2$ core-shell system in water, strong coupling regime can be achieved. Figure 2(d) shows the scattering cross section of the SiNP-WS$_2$ core-shell system in water as a function of wavelength and SiNP radius. These results confirm the expected anti-crossing behavior of scattering at the intersection of excitonic and MD resonances.

In order to experimentally demonstrate tunable resonance coupling for the single SiNP-monolayer WS$_2$ heterostructures, we have realized experimentally the geometry of optically resonant spherical SiNP arranged on a WS$_2$ monolayer schematically shown in Figure 3(a). We prepared a WS$_2$ monolayer by chemical vapor deposition (CVD) and then transferred it onto a glass substrate via PMMA-based transfer method (for details of monolayer WS$_2$ preparation and transfer see *Methods* and *Supplementary information*). Figure 3(b) shows an optical image of the WS$_2$ flake that we used. The monolayer nature of the WS$_2$ flake is confirmed by the strong PL signal (Figure 3(d)), which is typical for direct bandgap semiconductors[52]. The PL spectrum of monolayer WS$_2$ on the glass substrate in air under ambient conditions shows a resonance at 620 nm (1.99 eV), corresponding to A exciton, which matches the reported emission of CVD-grown monolayer WS$_2$. Moreover, the spectrum has a resonance at the negatively charged A$^-$ exciton (trion), which originates from slight electrical doping of the glass substrate. Atomic force microscopy (AFM) was also used to confirm the thickness of monolayered WS$_2$ (see *Supplementary information*). The crystalline structure of the monolayer WS$_2$ is also confirmed by Raman scattering (see Figure 3(c)). The Raman spectrum exhibits two peaks centered at 362 cm$^{-1}$ and 424 cm$^{-1}$, and that correspond to the out-of-plane E' mode, and the in-plane $A_1'$ mode of monolayer $WS_2$, respectively[51].

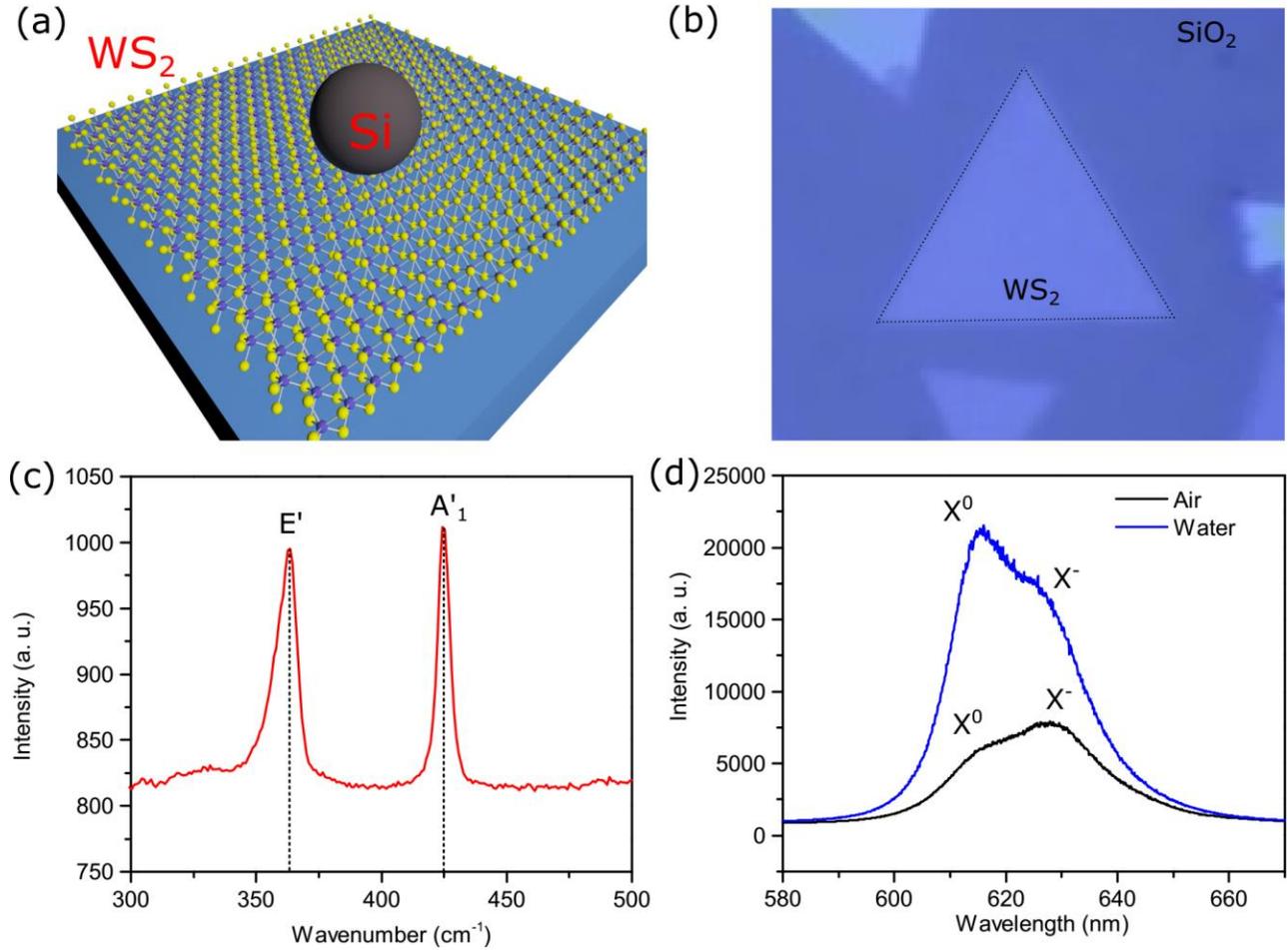

Figure 3: (a) Schematic view of a sample comprising a single SiNT on monolayer $WS_2$. (b) An optical image of monolayer $WS_2$. The scale bar is 20 μm. Raman (c) and photoluminescence (PL) (d) spectra of the monolayer $WS_2$ on $SiO_2$ in air (the excitation wavelength is 488 nm).

The PL spectra shown in Figure 3(d) reveals a 4 times enhancement in the PL intensity of the bare $WS_2$ monolayer in water. This allows us to estimate the change of dipole moment of a single exciton. To this end, we consider that the PL intensity $I$ is proportional to the squared total dipole $D$ moment of all excitons in the system $I \sim D^2 = (d \cdot N)^2$. In addition, the number of excited excitons $N$ is proportional to the tangential electric field intensity on the TMDC surface, which is reduced in water due to Fresnel reflection. Taking all of this into account, we find that the dipole moment of a single exciton in the WS$_2$ monolayer in water increases by ≈ 2.7 times compared to air. This value is not far from the estimation obtained above using the static screening approximation.

Hydrogenated amorphous silicon (a-Si:H) nanoparticles were synthesized in supercritical n-hexane (for details of a-Si:H NPs preparation see *Methods* and *Supplementary information*)[55]. The right inset in Figure 4(a) shows the SEM image of Si nanoparticle arranged on the WS$_2$ flake. Figure 4 shows the experimentally measured (blue line) and simulated by CMT (red dots) scattering spectra for single SiNP-monolayer WS$_2$ heterostructures in air (a) and water (b). In the case of air, we observe a Fano-like scattering feature at the wavelength of the A-exciton resonance. The fitting by CMT gives the Rabi splitting energy of 49.6 meV in this case. Based on the above, we can expect a significant enhancement of the Rabi splitting energy when the system is placed in water. Indeed, in this second scenario we achieve 86.6 meV. We emphasize that this experimentally measured value is much less than the one expected for the core-shell geometry, because of the significantly smaller number of excitons that take place in the resonance interaction. Considering that the SiNP has only one point of contact with the TMDC surface, we may estimate the number of excited excitons to be only a few.

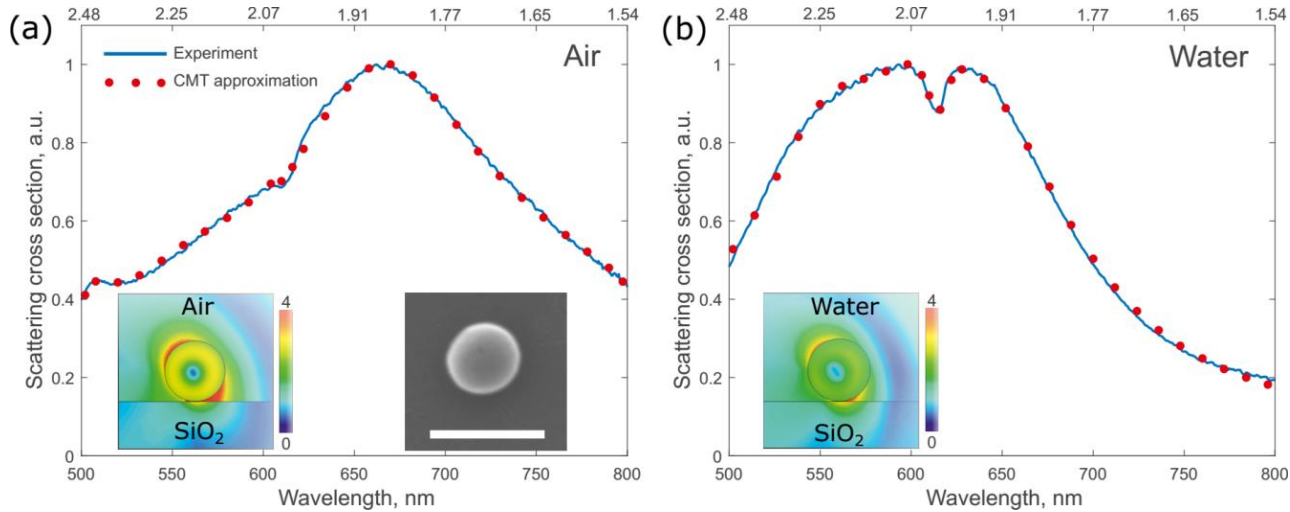

Figure 4: Experimentally measured (blue line) and fitted by CMT (red dots) scattering spectra for single SiNP-monolayer WS$_2$ heterostructures in air (a) and water (b). Left insets show distribution profiles of absolute values of electric field at the wavelength of excitonic resonance (620 nm), which is close to the magnetic resonance of SiNP. Right inset in (a)

shows SEM image of Si nanoparticle on the WS$_2$ flake. The scalebar is 250 nm. Values of Rabi splitting energy obtained from fitting of the experimental data demonstrate a significant enhancement from 49.6 meV up to 86.6 meV with replacing air by water.

Finally, the experimentally measured increase of the coupling constant, ~2 times, is in good agreement with the increase of the PL spectra for water by factor of 4, and the reduction of local electric field within SiNP and WS$_2$ flake by ~0.5 (see left insets in Figure 4). These distribution profiles show absolute values of the electric field at the wavelength of excitonic resonance (615 nm), which is close to the magnetic resonance of SiNP in air and water.

In conclusion, we have addressed the issue of resonance coupling in single Si nanoparticle-monolayer TMDC (WS$_2$) structures. We have theoretically predicted a transition from weak to strong coupling with a coupling constant exceeding 95 meV for a SiNP covered by monolayer WS$_2$ at the magnetic optical Mie resonance when changing the surrounding dielectric material from air to water. This prediction is based on the experimental estimation of the TMDC dipole moment change obtained from measured photoluminescence spectra of the TMDC in different solvents. The ability of such a system to realize tunable resonance coupling has been experimentally verified using optically resonant spherical Si nanoparticle placed on a WS$_2$ monolayer. The coupling constant obtained in this case increases from 24.8 meV up to 43.4 meV when replacing air by water as the background material. Our findings pave the way to a novel class of low-loss nanophotonic, optoelectronic and quantum optical devices based on dielectric resonant nanoparticles coupled with 2D excitonic materials.

## Methods

*Monolayer WS$_2$ preparation.* Atmospheric-pressure chemical vapor deposition (APCVD) method was used to prepare monolayer WS$_2$ on SiO$_2$/Si wafers. A tube furnace with Argon as the carrier gas (the flow rate is 150 sccm, during the synthesis) was employed. Two freshly cleaned SiO$_2$/Si wafers with ~10 mg WO$_3$ powders sandwiched by them were placed in a 2 cm-diameter quartz tube. Then, WO$_3$ powders was heated up to 700 °C and held for 15 min in the tube furnace, and simultaneously sulfur powders were separately heated up to 250 °C with a heating belt. Following the APCVD fabrication, the as-grown monolayer WS$_2$ was transferred onto a glass substrate. The transfer process began by spin coating a layer of PMMA on as-grown monolayer $WS_2$ on the $SiO_2$/Si wafer. Then, the $SiO_2$/Si wafer was etched away by 1 M sodium hydroxide (NaOH) aqueous solution. After being rinsed by DI-water for several times, the detached PMMA+$WS_2$ film was fished onto a pre-cleaned glass substrate. Finally, the PMMA layer was removed by immersing the whole specimen into acetone. For more details see *Supplementary information.*

*Synthesis of Si nanoparticles.* Hydrogenated amorphous silicon (a-Si:H) nanoparticles were synthesized in supercritical n-hexane, as described previously[56,57]. A 10 mL titanium batch reactor purchased from High Pressure Equipment Company (HiP Co.) was brought in a nitrogen filled glove box. 21 μL trisilane and a certain amount of n-hexane were added in the reactor. In all reactions, the pressure was kept at 34.5 MPa (5000 psi). After loading the reagents, the reactor was first sealed using a wrench inside the glove box. The reactor was then inserted into a heating block and heated to the target temperature for 10 min to allow complete decomposition of trisilane. After the reaction, the reactor was cooled to room temperature by an ice bath. The reactor was then opened and the colloidal a-Si:H nanoparticles were extracted from the reactor. The nanoparticles were washed using chloroform by centrifuging at 8000 rpm for 5 min. The precipitate was collected and dispersed in chloroform before use. For more details see *Supplementary information.*

*Optical measurements.* SiNPs were drop coasted on the bare glass substrate or the top of monolayer $WS_2$. The scattering spectra of single SiNPs on the bare glass substrate or monolayer $WS_2$ were measured by an inverted microscope (Ti-E, Nikon) with a spectrograph (Andor), an EMCCD (Andor) and a halogen white light source (12V, 100 W), as schematically shown in Figure S3 (see *Supplementary information*)[24,58,59]. To measure the scattering spectra in the solvent with a big dielectric constant, water was sandwiched between the sample and a cover glass. The Witec Micro-Raman Spectrometer was also used to measure PL spectra of monolayer $WS_2$ in different surrounding media.

## Author Contributions

S.L., A.K., M.W., Y.Z. and O.K. conceived the study and designed the experiments. S.L., A.K. and O.K. conducted the theoretical calculations. Y.Z., A.A., and M.W. participated in analysis of the theoretical and experimental data. T.J. and B.K. synthesized the silicon nanospheres. T.Z., H.L., and M.T. synthesized the WS$_2$ monolayers, conducted materials characterization and optimized the transferred process. O.K., A.K., A.A. and M.W. co-wrote the manuscript, with contributions of all authors. All authors have given approval to the final version of the manuscript.

$These authors contributed equally.

## Authors' information

The authors declare no competing financial interest.

## Acknowledgement

Y. Z. acknowledges the financial support of the Office of Naval Research Young Investigator Program (N00014-17-1-2424) and the National Science Foundation (NSF) Grant 1704634. M.T., T.Z and H.L thank NSF for support through the EFRI 2-DARE Grant 1542707. M. S. acknowledges the financial support of GS University Graduate Continuing Fellowship.

## References

(1)   Wang, Y.; Knoll, W.; Dostalek, J. *Anal. Chem.* **2012**, *84* (19), 8345–8350.

(2)   Butler, S. Z.; Hollen, S. M.; Cao, L.; Cui, Y.; Gupta, J. A.; Gutiérrez, H. R.; Heinz, T. F.; Hong, S. S.; Huang, J.; Ismach, A. F.; Johnston-Halperin, E.; Kuno, M.; Plashnitsa, V. V.; Robinson, R. D.; Ruoff, R. S.; Salahuddin, S.; Shan, J.; Shi, L.; Spencer, M. G.; Terrones, M.; Windl, W.; Goldberger, J. E. *ACS Nano* **2013**, *7* (4), 2898–2926.

(3)   Xia, F.; Wang, H.; Xiao, D.; Dubey, M.; Ramasubramaniam, A. *Nat. Photonics* **2014**, *8* (12), 899–907.

(4)   Mak, K. F.; Shan, J. *Nat. Photonics* **2016**, *10* (4), 216–226.

(5)   Bhimanapati, G. R.; Lin, Z.; Meunier, V.; Jung, Y.; Cha, J.; Das, S.; Xiao, D.; Son, Y.; Strano, M. S.; Cooper, V. R.; Liang, L.; Louie, S. G.; Ringe, E.; Zhou, W.; Kim, S. S.; Naik, R. R.; Sumpter, B. G.; Terrones, H.; Xia, F.; Wang, Y.; Zhu, J.; Akinwande, D.;

Alem, N.; Schuller, J. A.; Schaak, R. E.; Terrones, M.; Robinson, J. A. *ACS Nano* **2015**, *9* (12), 11509–11539.

(6) Das, S.; Robinson, J. A.; Dubey, M.; Terrones, H.; Terrones, M. *Annu. Rev. Mater. Res.* **2015**, *45* (1), 1–27.

(7) Zhao, W.; Ghorannevis, Z.; Chu, L.; Toh, M.; Kloc, C.; Tan, P. H.; Eda, G. *ACS Nano* **2013**, *7* (1), 791–797.

(8) Lopez-Sanchez, O.; Lembke, D.; Kayci, M.; Radenovic, A.; Kis, A. *Nat. Nanotechnol.* **2013**, *8* (7), 497–501.

(9) Britnell, L.; Ribeiro, R. M.; Eckmann, A.; Jalil, R.; Belle, B. D.; Mishchenko, A.; Kim, Y.-J.; Gorbachev, R. V.; Georgiou, T.; Morozov, S. V.; Grigorenko, A. N.; Geim, A. K.; Casiraghi, C.; Neto, A. H. C.; Novoselov, K. S. *Science (80-. ).* **2013**, *340* (6138), 1311–1314.

(10) Perea-Lõpez, N.; Elías, A. L.; Berkdemir, A.; Castro-Beltran, A.; Gutiérrez, H. R.; Feng, S.; Lv, R.; Hayashi, T.; Lõpez-Urías, F.; Ghosh, S.; Muchharla, B.; Talapatra, S.; Terrones, H.; Terrones, M. *Adv. Funct. Mater.* **2013**, *23* (44), 5511–5517.

(11) Koppens, F. H. L.; Mueller, T.; Avouris, P.; Ferrari, A. C.; Vitiello, M. S.; Polini, M. *Nat. Nanotechnol.* **2014**, *9* (10), 780–793.

(12) Yin, Z.; Li, H.; Li, H.; Jiang, L.; Shi, Y.; Sun, Y.; Lu, G.; Zhang, Q.; Chen, X.; Zhang, H. *ACS Nano* **2012**, *6* (1), 74–80.

(13) Sun, Z.; Martinez, A.; Wang, F. *Nat. Photonics* **2016**, *10* (4), 227–238.

(14) Lopez-Sanchez, O.; Alarcon Llado, E.; Koman, V.; Fontcuberta I Morral, A.; Radenovic, A.; Kis, A. *ACS Nano* **2014**, *8* (3), 3042–3048.

(15) Withers, F.; Del Pozo-Zamudio, O.; Mishchenko, A.; Rooney, A. P.; Gholinia, A.; Watanabe, K.; Taniguchi, T.; Haigh, S. J.; Geim, A. K.; Tartakovskii, A. I.; Novoselov, K. S. *Nat. Mater.* **2015**, *14* (3), 301–306.

(16) Liu, C. H.; Clark, G.; Fryett, T.; Wu, S.; Zheng, J.; Hatami, F.; Xu, X.; Majumdar, A. *Nano Lett.* **2017**, *17* (1), 200–205.

(17) Ye, Y.; Wong, Z. J.; Lu, X.; Ni, X.; Zhu, H.; Chen, X.; Wang, Y.; Zhang, X. *Nat. Photonics* **2015**, *9* (11), 733–737.

(18) Wu, S.; Buckley, S.; Schaibley, J. R.; Feng, L.; Yan, J.; Mandrus, D. G.; Hatami, F.; Yao, W.; Vučković, J.; Majumdar, A.; Xu, X. *Nature* **2015**, *520* (7545), 69–72.

(19) Wen, J.; Wang, H.; Wang, W.; Deng, Z.; Zhuang, C.; Zhang, Y.; Liu, F.; She, J.; Chen, J.; Chen, H.; Deng, S.; Xu, N. *Nano Lett.* **2017**, *17* (8), 4689–4697.

(20) Wang, S.; Li, S.; Chervy, T.; Shalabney, A.; Azzini, S.; Orgiu, E.; Hutchison, J. A.; Genet, C.; Samorì, P.; Ebbesen, T. W. *Nano Lett.* **2016**, *16* (7), 4368–4374.

(21) Zheng, D.; Zhang, S.; Deng, Q.; Kang, M.; Nordlander, P.; Xu, H. *Nano Lett.* **2017**, *17* (6), 3809–3814.

(22) Baranov, D. G.; Wersäll, M.; Cuadra, J.; Antosiewicz, T. J.; Shegai, T. *ACS Photonics* **2017**, *1*, acsphotonics.7b00674.

(23) Abid, I.; Chen, W.; Yuan, J.; Bohloul, A.; Najmaei, S.; Avendano, C.; Péchou, R.; Mlayah, A.; Lou, J. *ACS Photonics* **2017**, *4* (7), 1653–1660.

(24) Wang, M.; Li, W.; Scarabelli, L.; Rajeeva, B. B.; Terrones, M.; Liz-Marzán, L. M.; Akinwande, D.; Zheng, Y. *Nanoscale* **2017**, *9* (37), 13947–13955.

(25) Naik, G. V.; Shalaev, V. M.; Boltasseva, A. *Adv. Mater.* **2013**, *25* (24), 3264–3294.

(26) Khurgin, J. B. *Nat. Nanotechnol.* **2015**, *10* (1), 2–6.

(27) Kuznetsov, A. I.; Miroshnichenko, A. E.; Brongersma, M. L.; Kivshar, Y. S.; Luk'yanchuk, B. *Science (80-. ).* **2016**, *354* (6314), aag2472.

(28) Baranov, D. G.; Zuev, D. A.; Lepeshov, S. I.; Kotov, O. V.; Krasnok, A. E.; Evlyukhin, A. B.; Chichkov, B. N. *Optica* **2017**, *4* (7), 814.

(29) Evlyukhin, A. B.; Novikov, S. M.; Zywietz, U.; Eriksen, R. L.; Reinhardt, C.; Bozhevolnyi, S. I.; Chichkov, B. N. *Nano Lett.* **2012**, *12* (7), 3749–3755.

(30) Zywietz, U.; Schmidt, M. K.; Evlyukhin, A. B.; Reinhardt, C.; Aizpurua, J.; Chichkov, B. N. *ACS Photonics* **2015**, *2* (7), 913–920.

(31) Jahani, S.; Jacob, Z. *Nat. Nanotechnol.* **2016**, *11* (1), 23–36.

(32) Krasnok, A. E.; Miroshnichenko, A. E.; Belov, P. A.; Kivshar, Y. S. *Opt. Express* **2012**, *20* (18), 20599–20604.

(33) Li, Y.; Cain, J. D.; Hanson, E. D.; Murthy, A. A.; Hao, S.; Shi, F.; Li, Q.; Wolverton, C.; Chen, X.; Dravid, V. P. *Nano Lett.* **2016**, *16* (12), 7696–7702.

(34) Mi, Y.; Zhang, Z.; Zhao, L.; Zhang, S.; Chen, J.; Ji, Q.; Shi, J.; Zhou, X.; Wang, R.; Shi, J.; Du, W.; Wu, Z.; Qiu, X.; Zhang, Q.; Zhang, Y.; Liu, X. *Small* **2017**, *1701694*, 1701694.

(35) Bohren, Craig F, Huffman, D. R. *Absorption and Scattering of Light by Small Particles*; Bohren, C. F., Huffman, D. R., Eds.; Wiley-VCH Verlag GmbH: Weinheim, Germany, Germany, 1998.

(36) Li, Y.; Chernikov, A.; Zhang, X.; Rigosi, A.; Hill, H. M.; Van Der Zande, A. M.; Chenet, D. A.; Shih, E. M.; Hone, J.; Heinz, T. F. *Phys. Rev. B - Condens. Matter Mater. Phys.* **2014**, *90* (20), 1–6.

(37) Vuye, G.; Fisson, S.; Nguyen Van, V.; Wang, Y.; Rivory, J.; Abelès, F. *Thin Solid Films* **1993**, *233* (1–2), 166–170.

(38) Savelev, R. S.; Sergaeva, O. N.; Baranov, D. G.; Krasnok, A. E.; Alù, A. *Phys. Rev. B* **2017**, *95* (23), 235409.

(39) Alù, A.; Engheta, N. *J. Appl. Phys.* **2005**, *97* (9), 94310.

(40) Aden, A. L.; Kerker, M. *J. Appl. Phys.* **1951**, *22* (10), 1242–1246.

(41) Zengin, G.; Johansson, G.; Johansson, P.; Antosiewicz, T. J.; Käll, M.; Shegai, T. *Sci.*

Rep. **2013**, *3* (1), 3074.

(42) Wu, X.; Gray, S. K.; Pelton, M. *Opt. Express* **2010**, *18* (23), 23633.

(43) Lin, Y.; Ling, X.; Yu, L.; Huang, S.; Hsu, A. L.; Lee, Y. H.; Kong, J.; Dresselhaus, M. S.; Palacios, T. *Nano Lett.* **2014**, *14* (10), 5569–5576.

(44) Mao, N.; Chen, Y.; Liu, D.; Zhang, J.; Xie, L. *Small* **2013**, *9* (8), 1312–1315.

(45) Marquier, F.; Sauvan, C.; Greffet, J.-J. *ACS Photonics* **2017**, *4* (9), acsphotonics.7b00475.

(46) Słowik, K.; Filter, R.; Straubel, J.; Lederer, F.; Rockstuhl, C. *Phys. Rev. B - Condens. Matter Mater. Phys.* **2013**, *88* (19), 195414.

(47) Chikkaraddy, R.; de Nijs, B.; Benz, F.; Barrow, S. J.; Scherman, O. A.; Rosta, E.; Demetriadou, A.; Fox, P.; Hess, O.; Baumberg, J. J. *Nature* **2016**, *535* (7610), 127–130.

(48) Zhou, N.; Yuan, M.; Gao, Y.; Li, D.; Yang, D. *ACS Nano* **2016**, *10* (4), 4154–4163.

(49) Mahan, G. D. *Many-Particle Physics*; Springer US: Boston, MA, 2000.

(50) Stier, A. V.; Wilson, N. P.; Clark, G.; Xu, X.; Crooker, S. A. *Nano Lett.* **2016**, *16* (11), 7054–7060.

(51) Wang, G.; Chernikov, A.; Glazov, M. M.; Heinz, T. F.; Marie, X.; Amand, T.; Urbaszek, B. **2017**, No. Ml.

(52) Zhang, Y.; Zhang, Y.; Ji, Q.; Ju, J.; Yuan, H.; Shi, J.; Gao, T.; Ma, D.; Liu, M.; Chen, Y.; Song, X.; Hwang, H. Y.; Cui, Y.; Liu, Z. *ACS Nano* **2013**, *7* (10), 8963–8971.

(53) del Corro, E.; Botello-Méndez, A.; Gillet, Y.; Elias, A. L.; Terrones, H.; Feng, S.; Fantini, C.; Rhodes, D.; Pradhan, N.; Balicas, L.; Gonze, X.; Charlier, J.-C.; Terrones, M.; Pimenta, M. A. *Nano Lett.* **2016**, *16* (4), 2363–2368.

(54) Carvalho, B. R.; Wang, Y.; Mignuzzi, S.; Roy, D.; Terrones, M.; Fantini, C.; Crespi, V.

H.; Malard, L. M.; Pimenta, M. A. *Nat. Commun.* **2017**, *8*, 14670.

(55) Shi, L.; Harris, J. T.; Fenollosa, R.; Rodriguez, I.; Lu, X.; Korgel, B. A.; Meseguer, F. *Nat. Commun.* **2013**, *4* (May), 1904.

(56) Pell, L. E.; Schricker, A. D.; Mikulec, F. V.; Korgel, B. A. *Langmuir* **2004**, *20* (16), 6546–6548.

(57) Harris, J. T.; Hueso, J. L.; Korgel, B. A. *Chem. Mater.* **2010**, *22* (23), 6378–6383.

(58) Wang, M.; Hartmann, G.; Wu, Z.; Scarabelli, L.; Rajeeva, B. B.; Jarrett, J. W.; Perillo, E. P.; Dunn, A. K.; Liz-Marzán, L. M.; Hwang, G. S.; Zheng, Y. *Small* **2017**, *13* (38), 1701763.

(59) Wang, M.; Bangalore Rajeeva, B.; Scarabelli, L.; Perillo, E. P.; Dunn, A. K.; Liz-Marzán, L. M.; Zheng, Y. *J. Phys. Chem. C* **2016**, *120* (27), 14820–14827.